# The experimental determination of exchange mass terms in surface states on both terminations of MnBi$_4$Te$_7$

Dezhi Song[1†], Fuyang Hang[1†], Gang Yao[2], Jun Zhang[1]*, Ye-Ping Jiang[1]*, Jin-Feng Jia[3,4,5]

1 Key Laboratory of Polar Materials and Devices, Department of Electronic, East China Normal University, Shanghai 200241, China.

2 Chongqing key Laboratory of Micro & Nano Structure Optoelectronics, School of Physical Science and Technology, Southwest University, Chongqing 400715, China

3 Key Laboratory of Artificial Structures and Quantum Control (Ministry of Education), Tsung-Dao Lee Institute, School of Physics and Astronomy, Shanghai Jiao Tong University, Shanghai, 200240, China.

4 Southern University of Science and Technology, Shenzhen 518055, China.

5 Quantum Science Center of Guangdong-HongKong-Macao Greater Bay Area, Shenzhen 518045, China.

The intrinsic antiferromagnetic topological insulators in the Mn-Bi-Te family, composed of superlattice-like MnBi$_2$Te$_4$/(Bi$_2$Te$_3$)$_n$ ($n$ = 0, 1, 2, 3…) layered structure, present intriguing states of matter such as quantum anomalous Hall effect and the axion insulator. However, the surface state gap, which is the prerequisite for the observation of these states, remains elusive. Here by molecular beam epitaxy, we obtain two types of MnBi$_4$Te$_7$ films with the exclusive Bi$_2$Te$_3$ (BT) or MnBi$_2$Te$_4$ (MBT) terminations. By scanning tunneling spectroscopy, the mass terms in the surface states are identified on both surface terminations. Experimental results reveal the existence of a hybridization gap of approximately 23 meV in surface states on the BT termination. This gap comes from the hybridization between the surface states and the spin-split states in the adjacent MBT layer. On the MBT termination, an exchange mass term of about 30 meV in surface states is identified by taking magnetic-field-dependent Landau level spectra as well as theoretical simulations. In addition, the mass term varies with the field in the film with a heavy Bi$_{Mn}$ doping level in the Mn layers. These findings demonstrate the existence of mass terms in surface states on both types of terminations in our epitaxial MnBi$_4$Te$_7$ films investigated by local probes.

[†] These people contribute equally to this work.
* Corresponding authors. Email: zhangjun@ee.ecnu.edu.cn, ypjiang@clpm.ecnu.edu.cn

*Introduction.* The introduction of the exchange field into topological thin films may lead to the quantum anomalous Hall effect (QAHE) [1-4]. The first discovery of intrinsic magnetic topological insulators (MTI) in the Mn-Bi-Te family, composed of alternate stacking of the $MnBi_2Te_4$ (MBT) septuple layer (SL) and $(Bi_2Te_3)_n$ (BT) ($n$ = 0, 1, 2, 3…) quintuple layers (QL) [5-8], opens the possibility of realizing emergent phenomena such as QAHE[9,10] and axion insulator states[11,12] in MTI by the approach beyond magnetic doping. The A-type antiferromagnetic (AFM) ground states of these materials in this family, where the magnetic moments align ferromagnetically (FM) in the out-of-plane direction within each SL but align antiferromagnetically between SLs, make them more promising in the realization of above-mentioned exotic states. The AFM MTI changes into FM MTI by the spin-flop transitions at certain perpendicular magnetic fields, which decreases with the increasing BT intercalation $n$ between MBT SLs[13].

Although the zero-field QAHE has been realized in the cleaved samples of MBT crystals, the experimental investigation of the magnetic gap in the surface states, which constitutes the prerequisite condition to observe zero-field QAHE, remains controversial. These controversial experimental results mainly come from angle-resolved photoelectron spectroscopy (ARPES) measurements, where gapped[6,14] and gapless[15-17] surface states were both reported, with the latter contradicting theoretical predictions. Compared with ARPES, investigations made by the local probe, such as scanning tunneling spectroscopy (STS), revealed the simultaneous existence of gapped and gapless regions on MBT because of the inhomogeneity in the distribution of Mn concentration[18,19]. For $MnBi_4Te_7$ ($n$ = 1), recent ARPES measurements also illustrated the contentious surface state gap on both the MBT and BT terminations[20-38]. These underscore the complexity of the magnetic interactions and the significant role of defects and disorder in these materials.

We carried out the local investigation of surface states by STS on both terminations of $MnBi_4Te_7$ grown by molecular beam epitaxy (MBE). For the BT termination, a surface band structure with a well-defined energy gap is observed, in accord with the gapped feature resulting from the spin-orbital-coupled (SOC) overlapping bands from QL and SL[29,35]. On the MBT termination, Landau levels (LL) appear in the presence of a perpendicular magnetic field[39-41]. This enables the detailed investigation of surface state structures, where the combination of field-dependent LL spectra and numerical simulations indicates a mass term in surface states. The mass term yields a surface state gap of above 50 meV. Furthermore, the investigation of the MBT termination on the film with a moderate $Bi_{Mn}$ doping level implies a mass term that varies with the magnetic field in the surface states. Our local investigation validates the existence of a mass term in the surface states of $MnBi_4Te_7$ and demonstrates that this term can be tuned by non-magnetic doping in the ferromagnetic Mn layer. Our work establishes a framework for the fabrication of MTIs in the Mn-Bi-Te family with tunable magnetic structures.

*The growth*. Figure 1(a) demonstrates the side-view of the lattice structure of MnBi$_4$Te$_7$, which is formed by the alternate stacking of BT and MBT, with the layers bonded by van der Waals forces. Each MBT SL is composed of Mn, Bi, and Te layers stacked in the Te-Bi-Te-Mn-Te-Bi-Te sequence, where the Mn-on-Bi (Mn$_{Bi}$) and Bi-on-Mn (Bi$_{Mn}$) anti-site defects can be easily formed because of their relatively low formation energies[42]. This enables the epitaxial growth of MBT SLs with varying Bi$_{Mn}$ concentrations in the Mn layer by elaborating the Mn/Bi flux ratio during the film growth by MBE[18,43,44]. All experiments were carried out in a combined ultrahigh vacuum system composed of an MBE and a low-temperature scanning tunneling microscope (STM) equipped with a 14 T magnet. STM topography and spectroscopy were conducted at 4.3 K. The epitaxial MnBi$_4$Te$_7$ films were obtained by the repeated step-wise deposition of single QL of BT and SL of MBT on SrTiO$_3$ (STO) (111) substrates. The as-grown off-stoichiometric films have the form Mn$_{1-x}$Bi$_x$Bi$_{2(1-y)}$Mn$_{2y}$Te$_4$·Bi$_{2(1-z)}$Mn$_{2z}$Te$_3$. Here *x*, *y* and *z* denote the Bi$_{Mn}$ doping in the Mn layer (MBT), the Mn$_{Bi}$ doping in the Bi layer (MBT), and the Mn$_{Bi}$ doping in the Bi layer (BT), respectively.

Figure 1(b) illustrates the STM morphology of the MnBi$_4$Te$_7$ film composed of one QL BT and one SL MBT on the STO (111) substrate. As shown in the cross-sectional profile (Fig. 1c), the film has a uniform thickness of 2.40 ± 0.02 nm characteristic of MnBi$_4$Te$_7$ [35]. Here, the step-like features of ~ 0.24 nm in height come from the substrate. Figures 1(d) and 1(e) illustrate the Mn$_{Bi}$ defect (dark triangles) concentrations on the MBT (*y*) and BT (*z*) terminations, respectively[19,45]. Here the Mn$_{Bi}$ doping level *y* in the Bi layers of MBT seems to be independent of the Mn/Bi flux ratio and is fixed around 0.1. The Mn$_{Bi}$ doping level *z* in the Bi layers of BT is estimated to be around 0.015. Further increasing the Mn/Bi ratio during the deposition of BT does not lead to a higher doping level *z*. The concentration of Mn$_{Bi}$ defects in BT seems to be limited in the epitaxial MnBi$_4$Te$_7$ films.

*The mass term in surface states of MnBi$_4$Te$_7$ on the BT termination*. We measured the local density of states (LDOS) of the BT termination (BT/MBT/BT/MBT/BT) using *dI/dV* spectroscopy. Figures 2(a) and 2(b) show the spectra on a larger and a smaller energy scale, respectively. The larger-scale spectrum indicates a bulk-like gap of about 250 meV on the BT termination. The smaller-scale spectrum magnifies the LDOS coming from the surface states, where a 'U'-shaped gap-like feature near the Fermi level is clearly evident. To further characterize the nature of this gap, we investigated the magnetic field (out of plane) dependence of the surface state spectrum. As shown in Fig. 2(c), no obvious LL peaks can be observed and the energy gap does not change with the magnetic field. This result is in contrast with those of nonmagnetic TIs or magnetically doped TIs, where the LLs that evolve with the varying magnetic field can be clearly resolved due to the two-dimensional nature of surface states[40,41,46]. We exclude the defect scattering as one of the possible causes that suppresses the LLs. As will be shown later, LLs do appear instead on the MBT termination with a much larger $Mn_{Bi}$ concentration.

We attribute the indiscernible LLs on the BT termination to its anomalous surface state structure. According to the previous ARPES data and calculations, the surface states on the BT termination have a gap above the Dirac point due to their hybridization with the band coming from the underlying MBT layer. This leads to an energy gap of ~ 28 meV (Fig. 2(c)). Figure 2(d) shows the result of a simple modeling that demonstrates this hybridization effect (detailed in Note I of Supplementary Materials [47]). LLs with low indexes are more condensed near the gap edge because of the near-flat dispersion. The much smaller energy separations between these LLs make them more vulnerable to scattering and potential fluctuation. In addition, because of the hybridized nature, the spatial distribution of the resulting surface states is momentum dependent as illustrated in Fig. 2(d). For example, the states near $k = 0$ of the upper surface state branch mainly come from the underlying MBT layer.

Although the absence of LLs prevents the derivation of surface state dispersion, our modeling suggests that the presence of the gap-like feature does imply the existence of an exchange field (~ 28 meV) in the first BT and MBT layers. Meanwhile, as illustrated in Fig. S1, the hybridization gap near the Fermi level (Fig. 2(d)) only appears in the presence of the mass term (Fig. S1 of [47]). In fact, the gap comes from the hybridization between the Dirac conic surface state and the lower-lying spin-split band (caused by the exchange field). This indicates the presence of an out-of-plane net magnetic moment in the first MBT SL.

*The field-dependent LLs of surface states on the MBT termination.* Compared with the BT termination, magnetic-field dependent LLs can be observed on the MBT termination to enable the derivation of surface state structure. In this case, two types of samples with different $Bi_{Mn}$ defect concentrations ($x$) were prepared to investigate the interplay between the surface state structure and the magnetism. Figures 3(a) and 3(b) show the field-dependent LL spectra from 0 to 14 T on the MBT termination of samples with $x = 0$ and 0.3, respectively. There are two main features in these spectra on both terminations. The first is the appearance of a LL with an almost constant energy near where the surface LDOS approaches zero at zero field. The second is the appearance of two LLs above and below the LL mentioned above, with the former increasing and the latter decreasing in energy with the increasing magnetic field, respectively. We attribute the nearly constant LL and the other two LLs to the 0th, +1th and -1th LLs, respectively.

In addition, we see that the LLs with negative indexes are suppressed compared with the positive ones, especially for the $x = 0$ sample. This phenomenon can be explained by the difference in the Fermi velocities between upper and lower surface state branches as suggested by ARPES measurements [29,37]. The Fermi velocity of the surface states below the Dirac point decreases. This yields a near-flat band dispersion for the lower branch of surface states, especially in the presence of an exchange mass term. We model this behavior by considering the topological surface states with the Hamiltonian $H = \varepsilon_D + \frac{\hbar^2 k^2}{2m^*} + \hbar v_F \mathbf{k} \cdot \boldsymbol{\sigma} + m_z \sigma_z$, where $\varepsilon_D$, $m^*$, $v_F$, $\boldsymbol{\sigma}$ are the Dirac energy, the electron effective mass, the Fermi velocity and the spin Pauli matrices, respectively. Here we use the square term to describe the nonlinear Fermi velocity. The mass term $m_z$ is the exchange field along the $z$ direction. The resulting surface state structure $E = \varepsilon_D + \frac{\hbar^2 k^2}{2m^*} \pm \sqrt{\hbar^2 v_F^2 k^2 + m_z^2}$ with an energy gap $2m_z$ plotted in Fig. 3(c) reproduces the main features suggested by ARPES that differentiate the upper and lower surface state branches.

In the presence of a perpendicular magnetic field, the LLs are formed in the surface state spectrum. The Hamiltonian under a magnetic field can be written as $H = \varepsilon_D + \frac{\hbar^2 k^2}{2m^*} + \hbar v_F \mathbf{k} \cdot \boldsymbol{\sigma} + (m_z + \frac{1}{2} g_s \mu_B B)\sigma_z$, where is $\mathbf{k}$ the momentum after Peierls substitution ($\hbar \mathbf{k} \to \hbar \mathbf{k} - \mathbf{A}$) and $\frac{1}{2} g_s \mu_B B$ is the Zeeman term. The LL energies are: $E_n = \varepsilon_D + \frac{e\hbar B}{m^*}\left(n + \frac{1}{2}\right) \pm \sqrt{2e\hbar v_F^2 B|n| + \Delta^2}$ ($n = \pm 1, \pm 2, ...$) and $E_0 = \varepsilon_D + \frac{e\hbar B}{m^*} + \Delta$, where $\Delta = m_z + \frac{1}{2} g_s \mu_B B$ is the mass term. Neglecting the small Zeeman term gives $\Delta = m_z$. This yields the field-dependent LLs in Fig. 3(d), describing quite well the suppressed LLs with negative indexes.

Note that the -1th LL on the $x = 0.3$ sample is more apparent than that on the $x = 0$ sample. The above discussion yields one possible explanation for this behavior. In the heavily Bi$_{Mn}$ doped sample, the exchange term that mainly comes from the Mn layer may be reduced, leading to a smaller surface state gap and a less flat dispersion of the lower surface state branch as illustrated in Fig. 3(c), where the magnetic mass term $m_z$ of the surface state represented by the solid line is twice that of the dotted line. The field-dependent LLs are plotted in Fig. 3(d) for the case with larger $m_z$, where the -1the LL is nearly indistinguishable from other negative LLs. A smaller $m_z$ makes the dispersion of the lower surface state branch less flat. This makes the -1th LL more detached from other LLs and more discernible.

Another explanation is based on the fact that the surface state structure, especially the position of the Dirac point, changes with the Bi$_{Mn}$ doping. The electron doping level increases for the $x = 0.3$ film compared with the $x = 0$ film (Fig. S2 of [47]). This is consistent with the n-doping nature of the Bi$_{Mn}$ defects[19]. The position of the Dirac point, on the contrary, moves upward by $\sim$ 40 meV in the $x = 0.3$ film (Fig. 3), contradicting the heavier n-doping in this sample. Hence, the possible upward movement of the Dirac point makes the negative LLs with low indexes more detached from the valence band and more detectable.

*The derivation of the exchange mass term in surface states on the MBT termination.* Normally, from the field-dependent LL spectra, the surface state structure, including the exchange mass term, can be derived. Nonetheless, there is an anomaly in the experimental data compared with the simulated LLs as shown in Fig. 3(d) based on the above-mentioned surface state Hamiltonian. In contrast to the simulated one, the experimental 0th LL does not merge with +1th or -1th LLs at zero field according to its trend of evolution, as can be seen in Figs. 3(a) and 3(b). This controversy can be reconciled by adding a potential term into the Hamiltonian, describing the presence of potential fluctuations in our samples as indicated by the position-dependent LL spectra (Fig. S4 of [47]). In this case, the LLs near positions having local potential extremums are strongest, away from which the LL width increases because of splitting[48,49]. The field-dependent LLs in Fig. 3 and in Fig. S3 of [47] were taken on positions with the strongest LL features. Here, a parabolic potential $U(r) = \kappa r^2$ is added to the Hamiltonian for this purpose (Fig. 4(c)).

Figure 4(a) shows the simulated evolution of LLs along with the 0th and 1th LLs on the MBT termination of the $x = 0$ film (see the data taken on another two positions in Fig. S5(a) of [47]). A mass term $m_z$ of 30 meV and an appropriate $\kappa$ are used to get the best fitting (detailed in Note II of [47]). In the presence of the potential $U(r)$, the simulation indicates that the 0th LL is nearly constant and the 1th LL is almost linear with respect to the field in the higher-field region. The 0th LL is pushed up at lower fields by the potential as indicated in Fig. 4(c) and approaches the band edge at higher fields. For 1th LL, the influence of the potential extends to much higher fields because of the more extended wavefunctions of non-zeroth LLs, leading to the discrepancy between the extrapolation of the 0th LL and those of non-zeroth ones at zero field.

For the $x = 0.3$ film, the 0th LL at higher fields does not approach a constant value (see the data taken on another two positions in Fig. S5(b) of [47]). Note that the 0th LL indicates the band edge of the upper branch of surface states as well as the exchange mass $m_z$. Thus, the varying 0th LL in this case indicates the varying $m_z$ in response to the magnetic field. Figure 4(b) shows the simulation that takes into account a linear dependence of $m_z$ on the magnetic field.

Figure 4(d) shows the field-dependent exchange mass $m_z$ both for the nominally undoped and the heavily $Bi_{Mn}$ doped films. Assuming that $m_z$ mainly comes from the magnetic moments of Mn-layer spins, we can see that in the presence of heavy non-magnetic doping, the Mn-layer moments do not saturate at a moderate field as those of the undoped film do [50].

*Conclusions.* We have investigated the topological surface states on the two terminations of $MnBi_4Te_7$ using the local probes such as LL spectroscopy. By comparing the experimental data with both analytical and numerical simulations, the exchange term $m_z$ can be derived for the surface states on both terminations. On the BT termination, this exchange term leads to a surface state gap of about $m_z$ (~ 23 meV). On the MBT termination, this term (~ 30 meV) leads to a gap of about $2m_z$. The exchange mass terms obtained on both terminations are comparable, indicating that the surface states feel the exchange field mainly coming from the first MBT SL. In addition, on the MBT termination, two films with different $Bi_{Mn}$ dopings in the Mn layer were investigated, where the field-dependent $m_z$ was obtained by LL spectroscopy. We find that in the film with nominally zero doping, the exchange term $m_z$ saturates at a relatively low magnetic field, and that in the film with heavy doping, this term does not saturate even at relatively high magnetic fields. Our findings validate the existence of the exchange mass term on both terminations in epitaxial $MnBi_4Te_7$ films and show that this term can be tuned by non-magnetic doping, providing a possibility of manipulation of the interplay between magnetism and topology.

**Acknowledgments**

We acknowledge the supporting from National Key R&D Program of China (Grants No. 2022YFA1403102) and National Science Foundation (Grants No. 12474478, 92065102, 61804056).

**Supplementary Materials**

Figs. S1 to S5

**Figure Captions**

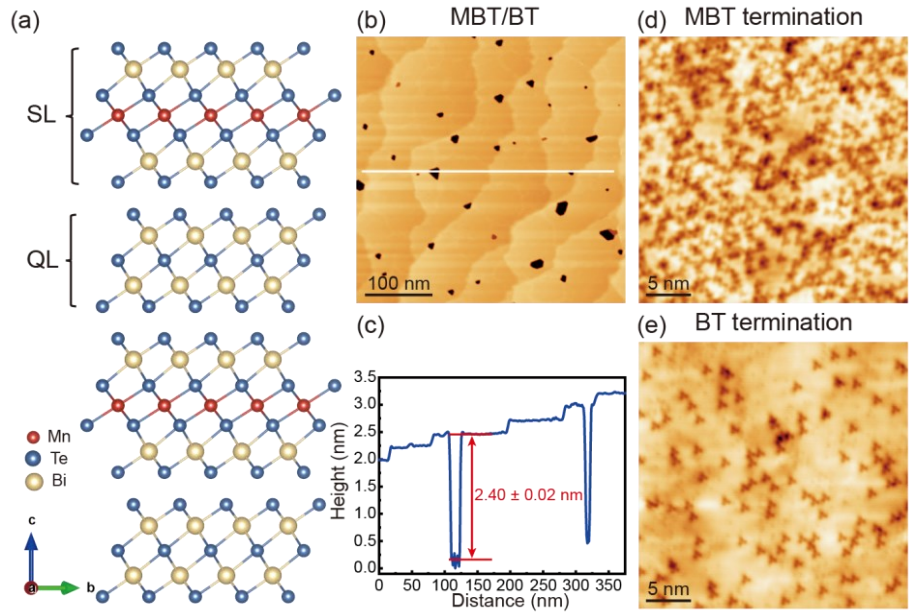

Fig. 1. The structure and the STM topography of epitaxial MnBi$_4$Te$_7$ films. (a) Schematic of the lattice structure of MnBi$_4$Te$_7$. (b) Large-scale STM image (3 V, 10 pA) of the MBT/BT film grown on STO (111). (c) The height profile of the film along the line in (b). (d) and (e) Typical STM images (1 V, 50 pA) on the MBT and BT terminations, respectively.

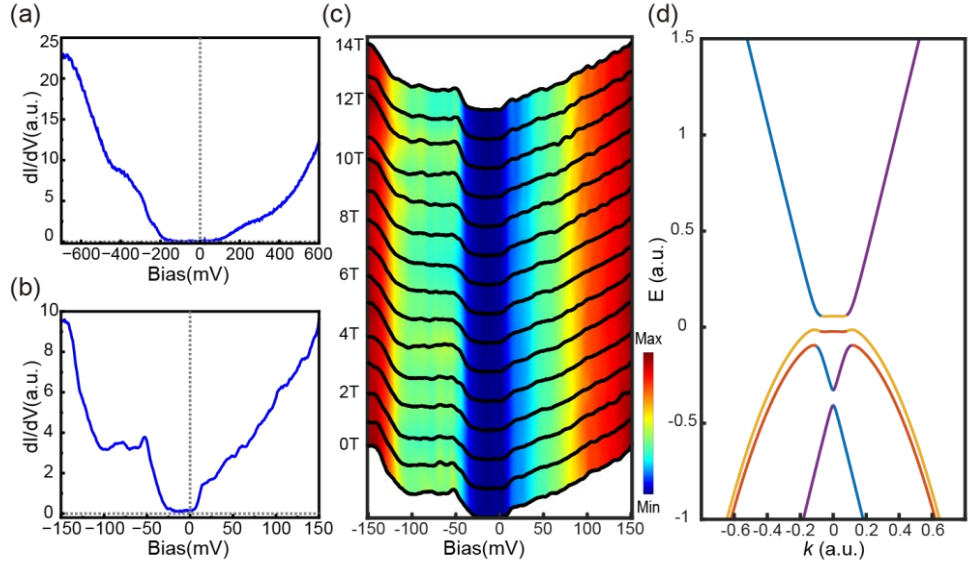

Fig. 2. The surface states on the BT termination (BT/MBT/BT/MBT/BT). (a) and (b) Characteristic STS on the BT termination in the larger and smaller energy scales, respectively. (c) STS in the magnetic fields from 0 to 14 T. (d) Simulation of the hybridization gap in the surface states.

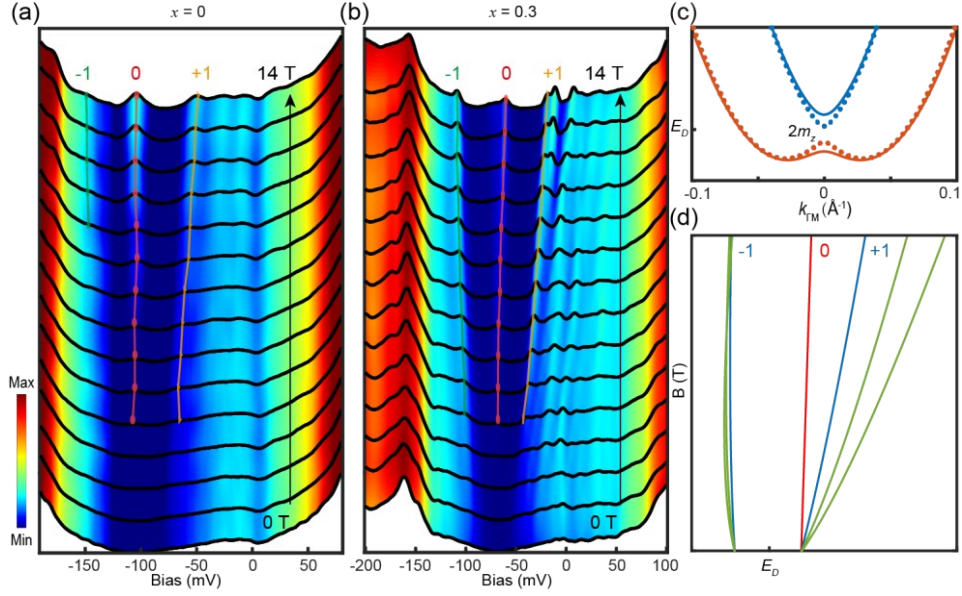

Fig. 3. Field-dependent Landau level spectra on the MBT termination for differently doped MnBi$_4$Te$_7$ films (MBT/BT/MBT/BT/MBT/BT). (a) and (d) Experimentally field-dependent Landau level spectra for the $x = 0$ and $x = 0.3$ films, respectively. (c) Simulation of the surface state structures on the MBT termination with different exchange mass $m_z$, where the dashed one has a smaller $m_z$. (d) The corresponding Landau level spectra (calculated) for the surface states with the larger $m_z$ in (c).

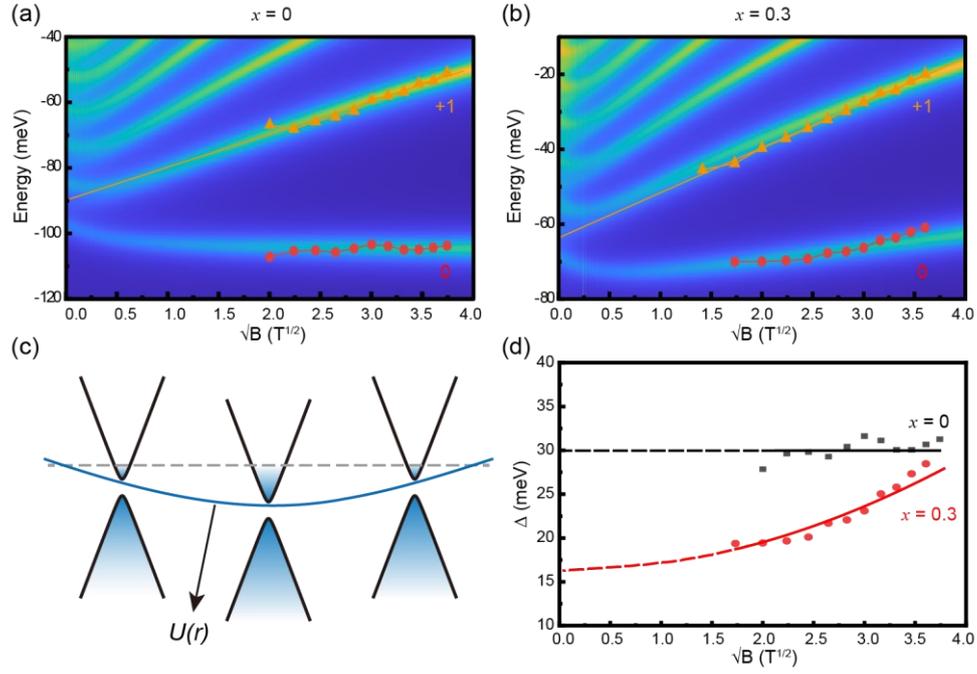

Fig. 4. Determination of the exchange term on the MBT termination. (a) and (b) Experimental LLs along with the calculated ones for the $x = 0$ and $x = 0.3$ films, respectively where the yellow triangles and red dots are the +1th and 0th LLs, respectively. (c) Schematic of a potential minimum modeled by $U(r)$ in the presence of potential fluctuation. (d) The derived field-dependent exchange mass term $m_z$ for different doped films.

# Supplementary Materials for

## The experimental determination of exchange mass terms in surface states on both terminations of MnBi$_4$Te$_7$

Dezhi Song[1†], Fuyang Hang[1†], Gang Yao[2], Jun Zhang[1]*, Ye-Ping Jiang[1]*, Jin-Feng Jia[3,4,5]

1 Key Laboratory of Polar Materials and Devices, Department of Electronic, East China Normal University, Shanghai 200241, China.
2 Chongqing key Laboratory of Micro & Nano Structure Optoelectronics, School of Physical Science and Technology, Southwest University, Chongqing 400715, China
3 Key Laboratory of Artificial Structures and Quantum Control (Ministry of Education), Tsung-Dao Lee Institute, School of Physics and Astronomy, Shanghai Jiao Tong University, Shanghai, 200240, China.
4 Southern University of Science and Technology, Shenzhen 518055, China.
5 Quantum Science Center of Guangdong-HongKong-Macao Greater Bay Area, Shenzhen 518045, China.

CONTENTS:
Note I to II
Figs. S1 to S5

[†] These people contribute equally to this work.
* Corresponding authors. Email: zhangjun@ee.ecnu.edu.cn, ypjiang@clpm.ecnu.edu.cn

Note I: The simulation of the hybridization gap.

In this section, we reproduce the calculations in the article[1] about the band structure that opens the gap induced by the hybridization effect between the top two layers[2,3]. This hybridization gap can be simply understood in terms of the gap arising from the coupling of the top and bottom surfaces of a 3D topological insulator in the 2D limit[4]. Following a standard $k \cdot p$ model, the band hybridization between QL and SL effective Hamiltonian reads:

$$H_s = \begin{pmatrix} m_z & iv_F k_x - vk_y & V_h & 0 \\ -iv_F k_x - vk_y & -m_z & 0 & V_h \\ V_h & 0 & \frac{\hbar^2 k^2}{2m^*} + m_z + \epsilon_D & 0 \\ 0 & V_h & 0 & \frac{\hbar^2 k^2}{2m^*} - m_z + \epsilon_D \end{pmatrix}$$

Here, $m^*$ is the electron effective mass of the electron in the SL, $\epsilon_D$ is the Dirac energy measured from the SL band, and $v_F$ is the Fermi velocity of QL. The $m_z$ represents the magnetic exchange interactions along the z direction, while the $V_h$ is a parameter describing the hybridization strength between QL and SL. When $m_z = 0$ and $V_h = 0$, a linear Dirac band from QL (purple and blue colors indicate spin-up and spin-down, respectively) and a quadratic band from SL on the surface. With the addition of the magnetic interaction ($m_z = 0.04$, $V_h = 0$), the Dirac band opens the mass gap, while the quadratic band shows Zeeman splitting properties (yellow and orange colors indicate spin-up and down properties, respectively). Finally, as illustrated in Fig. 2(d), due to the hybridization effect between QL and SL ($m_z = 0.04$, $V_h = 0.08$), the band dispersion displays a noticeable hybridization gap and inversion features. However, if only the hybridization effect between QL and SL is present ($m_z = 0$, $V_h = 0.08$), there is no significant hybridization gap observable.

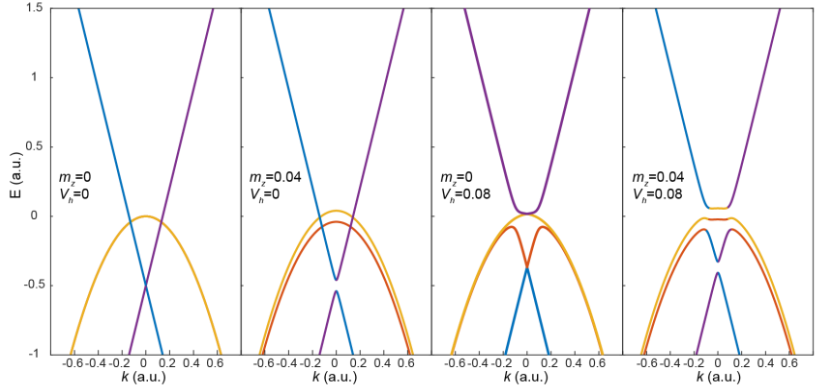

Fig. S1. Surface energy band structures that change with hybridization and magnetic exchange interactions

Note II: The simulation of the Landau level in the potential field.

To supplement the dependence of the Landau level on the magnetic field, we use an effective Hamiltonian similar to the article[5,6].

Here, $r$ and $B$ are simplified by $R_0$ and $B_*$, the length scale $R_0 = 60$ nm, and the scale for the magnetic field is thus $B_* = \hbar/(R_0^2 e) \sim 0.183$ T. $\varepsilon$ and $\Delta$ are simplified by the energy scale $\varepsilon_* = \hbar v_F/R_0 \sim 4.72$meV. We use $\kappa = 25$meV/nm$^2$. We use a Fermi velocity $v_F$ of $\sim 5.4\times10^5$ m/s$^2$ for x=0 according to the literature[7], and the $\Delta$ for Fig. 4(a) is 30meV, Fig. S5(a) left is 26meV and Fig. S5(a) right is 28meV.

We use a Fermi velocity $v_F$ of $\sim 4.7\times10^5$ m/s$^2$ for x=0.3, and the $\Delta$ for Fig. 4(a) is 15-28meV, Fig. S5(a) left is 12-28meV and Fig. S5(a) right is 10-28meV. Here the different Fermi velocities arise from the different energy band structures due to doping.

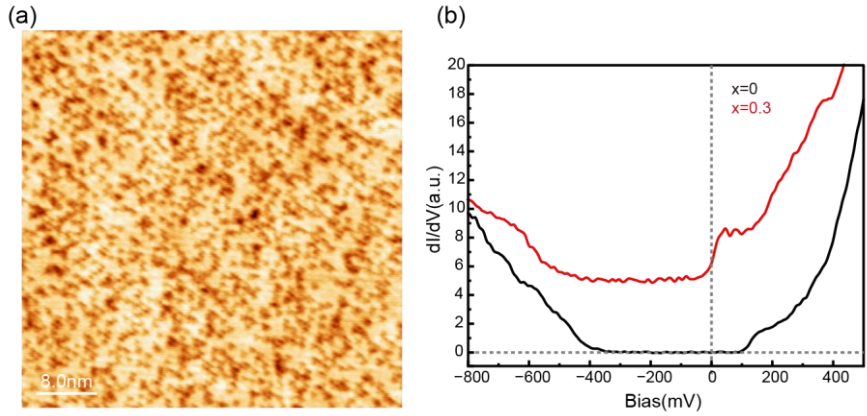

Fig. S2. (a) STM of surface defect distribution at x = 0.3 doping. (b) The dI/dV spectra taken on MBT/BT films with different Bi$_{Mn}$ doping.

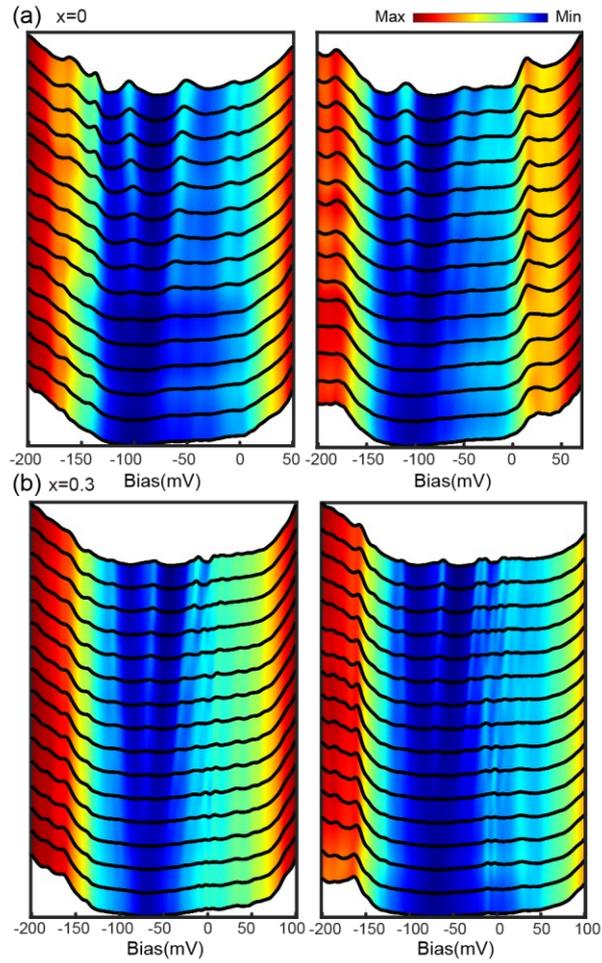

Fig. S3. (a) Field-dependent Landau level spectra of intrinsically doped MBT terminations obtained at different points in the experiment. (b) Field-dependent Landau level spectra of x = 0.3 doped MBT terminations obtained at different points in the experiment.

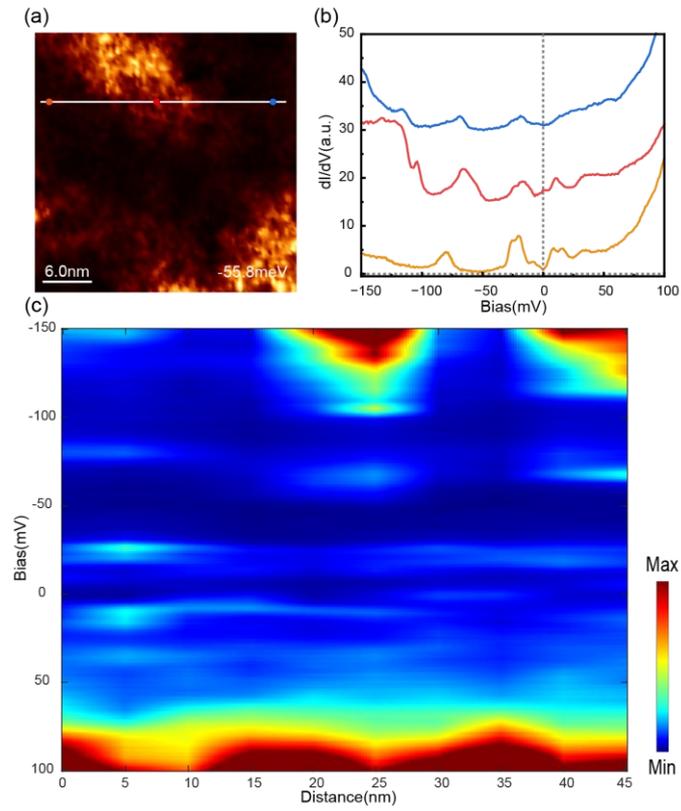

Fig. S4. (a) STS mapping at −55.8 meV showing the existence of doping variations in different regions. (b) The point spectra taken at the orange, red, and blue dots in (a). (c) The position-dependent dI/dV spectra along the white line in (a).

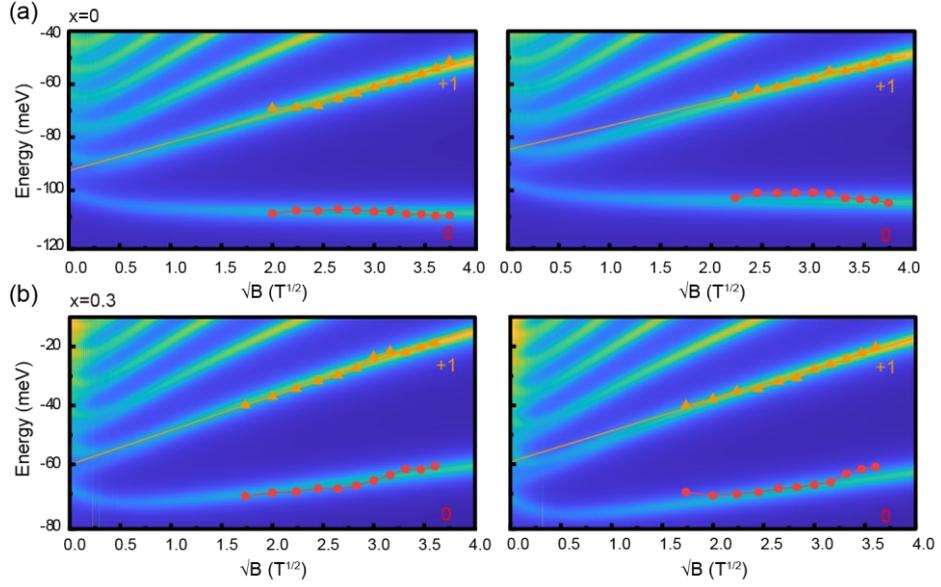

Fig. S5. (a) Experimental versus calculated fits for the zeroth LL and +1 LL at intrinsic doping corresponding to Fig. S2 (a). (b) The experimental versus calculated fits for the zeroth LL and +1 LL at x = 0.3 doping, which corresponds to Fig. S2 (b).